\documentclass[titlepage,12pt]{utarticle}
\usepackage{graphicx}
\usepackage{bm}
\usepackage{epsfig}
\usepackage{color}
\usepackage{colordvi}
\usepackage{amsmath,amssymb}
\usepackage{epsfig}
\usepackage{mathrsfs}

\newcommand{\beq}{\begin{equation}}
\newcommand{\eeq}{\end{equation}}
\newcommand{\bea}{\begin{eqnarray}}
\newcommand{\eea}{\end{eqnarray}}
\newcommand{\phib}{\overline{\varphi}}
\newcommand{\dphi}{\delta\varphi}
\newcommand{\ddphi}{\delta\dot{\varphi}}
\newcommand{\dddphi}{\delta\ddot{\varphi}}

\numberwithin{equation}{section}

\begin{document}

\title{Delayed Reheating and the Breakdown\\[.3cm] of Coherent Oscillations}
 
\author{Richard Easther, Raphael Flauger, and James B. Gilmore
   \address{
      Department of Physics,\\
      Yale University,\\
      New Haven, CT 06520, USA\\
      {~}\\[.5cm]
      {\rm {\it e}-mail}\\[.1cm]
      \emailt{richard.easther@yale.edu}\\
      \emailt{raphael.flauger@yale.edu}\\
      \emailt{james.gilmore@yale.edu}\\
   }
}

\Abstract{  We  analyze the evolution of the perturbations in the inflaton field and metric following the end of inflation. We present accurate analytic approximations for the perturbations, showing that the coherent oscillations of the post-inflationary condensate necessarily break down long before any current phenomenological constraints require the universe to become radiation dominated. Further, the breakdown occurs on length-scales equivalent to the comoving post-inflationary horizon size.   This work has implications for both the inflationary ``matching'' problem, and the possible generation of a stochastic gravitational wave background in the post-inflationary universe.\vskip-.8cm}

\maketitle

\section{Introduction}

A complete inflationary scenario must provide a
``graceful exit'' from the phase of accelerated expansion and (re)-thermalize the universe, setting the stage for a hot big bang.
In many models this process is efficiently driven by parametric
resonance sourced by the oscillating inflaton field~\cite{Traschen:1990sw,Kofman:1994rk,Kofman:1997yn}. The rich
phenomenology of preheating is often contrasted with ``standard'' reheating, where the universe is rethermalized via the decay of the inflaton into standard matter. Given that the couplings responsible for the decay must be small in order to protect the inflaton potential against loop corrections,
reheating is assumed to be a slow and gradual process
(see e.g.~\cite{Albrecht:1982mp}).
Without parametric resonance, it might appear that reheating could take place almost arbitrarily slowly: we must only {\em insist\/} that the universe reheat to MeV scale temperatures, a scale about eighteen orders of magnitude below the characteristic energy at the end of GUT scale inflation, in order to be
rethermalized in time for Big Bang Nucleosynthesis and to
generate a cosmological neutrino background~\cite{Komatsu:2010fb}.

In standard single field scenarios the inflaton rolls slowly through the inflationary portion of the potential,  and then performs rapid, damped oscillations about a minimum of the potential after inflation is over. The frequency of these oscillations is much higher than the expansion rate of the universe, so $\rho$ is almost constant during a single oscillation. Conversely, the pressure $p$ fluctuates rapidly: at maximum amplitude $\dot\varphi =0$  and $\rho= -p = V(\varphi_{max})$ while at the bottom of the potential $V(\varphi)=0$  and $\rho = p = \dot\varphi^2/2$. The equation of state parameter, $w$, thus varies between $\pm1$ during the oscillatory phase. Eventually, the amplitude of these oscillations will be small enough so that the potential is well approximated by the first term in its Taylor expansion around the minimum, $i.e$ a quadratic potential. The {\em average\/} equation of state can then be shown to be zero.  This leads to the conclusion that in the post-inflationary universe the energy density decays like $1/a^3$, and the scale factor grows like $a(t) \propto t^{2/3}$, mimicking a matter dominated universe.
Since the energy density can decrease by 72 orders of magnitude before reheating must occur, the universe can expand by a factor of up to $10^{24}$ during this effectively matter dominated period.\footnote{This is  a conservative limit but the precise scale one adopts here has no impact on our conclusions.}

In this paper, we carefully analyze the evolution of the perturbations in both the scalar field and the metric during this period of coherent oscillations.
Given that there are two well-separated time scales in this system (the Hubble time and the oscillation period), we can obtain accurate analytic approximations  for the field evolution by expanding in the ratio of the time scales. In the limit that the inflaton is coupled {\em only\/} to gravity and has a purely quadratic potential, we show that the perturbations grow and the system becomes nonlinear after the scale factor $a(t)$ has increased by a factor of about $10^6$. We expect that keeping higher order terms in the Taylor series expansion of the potential about its minimum lead to an even earlier onset of non-linearities, making this a conservative estimate. Consequently, while we are free to assume that the universe is matter dominated from the end of inflation to the MeV scale, the phase of coherent oscillations is necessarily much shorter.
We see that the perturbations that become non-linear first are those for which the physical momentum becomes comparable to the expansion rate of the universe at the end of inflation. The coherent oscillations thus fragment on length scales roughly equal to the (comoving) horizon size at the end of inflation.
 
We can understand our results heuristically by noting that in a universe dominated by pressureless dust, sub-horizon perturbations  grow linearly with  $a(t)$.   Given the apparent red tilt of the primordial perturbation spectrum, perturbations on scales near the post-inflationary horizon volume have a lower initial amplitude than those at longer scales, but they also spend more time inside the horizon during this effective matter dominated era. The growth of these modes during this period more than compensates for the diminished initial amplitude, and shorter modes become nonlinear before longer modes. However, modes whose frequency is higher than the oscillation frequency of the coherent field  ``resolve'' the oscillations. At this point our the analogy with a simple $p=0$ fluid  breaks down, and modes with higher frequencies will be suppressed.

Given the age and the simplicity of this model, these issues have
been touched upon a number of times in the past. Nambu and Sasaki consider a related problem in the context of the invisible axion \cite{Nambu:1989kh}, while Nambu and Taruya \cite{Nambu:1996gf} write down the equations of motion for the inflaton perturbations and scalar metric fluctuations, showing that they are described by a Mathieu-like equation.
Leach and Liddle \cite{Leach:2000yw} compute the spectrum of
perturbations for modes that leave the horizon near the end of
inflation,  while  Assadullahi and Wands
\cite{Assadullahi:2009nf,Assadullahi:2009jc} show that
nonlinearities formed during a generic matter dominated phase
likely leads to gravitational wave production, and  a
high-frequency, stochastic  background of gravitational waves in the
present-day universe. Recently,   Jedamzik, Lemoine and Martin
\cite{Jedamzik:2010dq} (see also \cite{Jedamzik:2010hq}) discussed
the gravitational growth of perturbations in this system, highlighting
 the  parametric resonance ``hidden'' in the scalar field
perturbations (see also \cite{Nambu:1996gf}). 
 
The purpose of this analysis is to understand and clarify the connection between the detailed evolution of inflaton and metric perturbations.  In particular, we carefully explore the  basis of the simple analogy between the post-inflationary coherent oscillations and a dust-dominated universe. We show that the universe necessarily  becomes inhomogeneous (and the perturbations nonlinear) if reheating is delayed long enough, and thus compute the maximum duration of a phase of coherent oscillations.  We are careful to perform our post-inflationary analysis using quantities that are well-defined at all times: some perturbation  variables which are well-defined during inflation have the background field velocity $\dot{\varphi}$ in the denominator, and thus become briefly singular during each oscillation of the field. The use of variables that remain smooth is particularly helpful for the numerical calculations we use to verify our analytic results.

The layout of this paper is as follows. In Section 2, we briefly review linear perturbation theory for a single scalar field with canonical kinetic term minimally coupled to gravity. We focus on the appropriate choice of variables for this analysis, since the standard Mukhanov-Sasaki variable exhibits singularities for modes
inside the horizon during the oscillatory phase. In Section 3, we study the evolution of perturbations that are near-horizon sized at the
end of inflation for a quadratic potential. We calculate an appropriately defined power spectrum, recovering the results by Leach and Liddle \cite{Leach:2000yw} where our calculation overlaps with theirs. Even though the underlying equation of state is rapidly oscillating, we find that the {\em average\/} growth of the density contrast of the modes is linear in the scale factor once they re-enter the horizon just like one would expect for a matter dominated universe.
Finally, we compute when the first nonlinearities
appear, verifying the estimate given above. We see that
the first nonlinearities are found on scales comparable to the
horizon size at the end of inflation. In Section 4, we discuss our
results and highlight a number of open questions identified by our
analysis.


\section{General Preliminaries}

We consider single field models of inflation in which the inflaton has a canonical kinetic term and is minimally coupled to gravity.  We first review the description and evolution of the perturbations in order to identify the most appropriate variables (see \cite{Weinberg:2008zzc} for a thorough introduction to cosmological perturbation theory).  In our  conventions the action is
\begin{equation}
S=\frac{1}{16\pi G}\int\;d^4x\sqrt{-g}R+\int\;d^4x\sqrt{-g}\left[-\frac12g^{\mu\nu}\partial_\mu\varphi\partial_\nu\varphi-V(\varphi)\right]\,.
\end{equation}
Working to first order in the perturbations, we make the usual decomposition into a position independent background and inhomogeneous perturbations:
\begin{equation}
\varphi(t,{\bf x})=\phib(t)+\dphi(t,{\bf x})\,.
\end{equation}
Ignoring tensor modes, we fix the gauge so that the  line element is
\begin{equation}
ds^2=-N^2dt^2+a^2(t)\delta_{ij}(dx^i+N^idt)(dx^j+N^jdt)\,.
\end{equation}
The  background geometry is governed by the usual Friedmann equation
\begin{equation}
\left(\frac{\dot{a}}{a}\right)^2=\frac{8\pi G}{3}\left(\frac12\dot\phib^2+V(\phib)\right)\,,\\
\end{equation}
while the homogeneous scalar field obeys
\begin{equation} \label{eq:phibackground}
\ddot\phib+3H\dot\phib+V'(\phib)=0\,,
\end{equation}
where $H=\dot{a}/{a}$ is the Hubble parameter and a prime indicates a derivative with respect to the field.

The quadratic part of the action will govern the evolution of the perturbations. As usual, we use the Hamiltonian and momentum constraints to express the perturbations in the lapse and the shift, $\delta N\equiv N-1$ and $N^i$, in terms of the scalar field perturbations. At first order, the momentum constraint implies
\begin{equation}
\delta N=\frac{4\pi G \dot\phib}{H}\dphi\,.
\end{equation}
Taking this result and the Hamiltonian constraint we see
\begin{equation}\label{eq:consni}
\partial_i N^i=-4\pi G\frac{\dot\phib^2}{H^2}\frac{d}{dt}\left(\frac{H}{\dot\phib}\dphi\right)\,.
\end{equation}
Using these expressions for $\delta N$ and $N^i$, the background  equations of motion, and integrating by parts, the quadratic part of the action for the perturbations becomes
\begin{eqnarray}
&&\nonumber\hskip-1.5cm S=\frac12\int\;dtd^3x\;a^3(t)\left[\ddphi^2-\frac{1}{a^2}(\partial\dphi)^2-V''(\phib)\dphi^2\right.\\
&&\hskip 3.25cm\left.+2\epsilon\eta H^2\dphi^2+6\epsilon H^2\dphi^2-2\epsilon^2H^2\dphi^2\vphantom{\frac{1}{a^2}}\right]\,,
\end{eqnarray}
where
\begin{equation}
\epsilon\equiv-\frac{\dot{H}}{H^2}=4\pi G\frac{\dot\phib^2}{H^2}\;\;\;\;\text{and}\;\;\;\;\eta\equiv\frac{\dot{\epsilon}}{\epsilon H}\,,
\end{equation}
are the Hubble slow-roll parameters.

As long as $\dot\phib\neq 0$ -- which is certainly true {\em during\/} inflation -- we can eliminate $V''(\phib)$ using the equation of motion. To be more specific,  one can differentiate equation~(\ref{eq:phibackground}) with respect to time, and obtain an expression for $\dot\phib V''(\phib)$. Provided that $\dot\phib\neq 0$, we can divide by $\dot\phib$, integrate by parts, and drop the boundary term to obtain the action
\begin{equation}
S=\frac12\int\;dtd^3x\; a^3(t)\left[\left(\ddphi-\frac12\eta H\dphi\right)^2-\frac{1}{a^2}(\partial\dphi)^2\right]\,.
\end{equation}
Varying this action with respect to $\dphi$ yields the Mukhanov-Sasaki equation, which can be brought into its standard form by making the identification $\dphi=v/a$, and using conformal time $\tau\equiv\int^t dt'/a(t')$ as the independent variable.

Here we are interested in the transition between inflation and non-accelerated expansion, and the post-inflationary field oscillations.  At the turning points of these oscillations, $\dot\phib=0$ and consequently $\epsilon=0$, so that $\eta$ becomes singular. These singularities arise because the boundary term that was dropped in eliminating $V''(\phib)$ becomes infinite when $\dot\phib=0$. We thus choose not to eliminate $V''(\phib)$ from the action, and instead express the slow-roll parameters in terms of the potential and its derivatives. Using the identities
\begin{equation}
(3-\epsilon)H^2=8\pi G V(\phib)\,,
\end{equation}
and
\begin{equation}
(6+\eta-2\epsilon)\epsilon H^2=-8\pi G V'(\phib)\frac{\dot\phib}{H}\,,
\end{equation}
the action becomes
\begin{eqnarray}
&&\nonumber\hskip-1.5cm S=\frac12\int\;dtd^3x\; a^3(t)\left[\vphantom{\frac{\dot\phib^2}{H^2}}\ddphi^2-\frac{1}{a^2}(\partial\dphi)^2-V''(\phib)\dphi^2\right.\\
&&\hskip3.25cm\left.-16\pi G\frac{\dot\phib}{H}V'(\phib)\dphi^2-(8\pi G)^2\frac{\dot\phib^2}{H^2}V(\phib)\dphi^2\right]\,.
\end{eqnarray}
The equation of motion for the perturbations in the scalar field is then
\begin{equation}\label{eq:eomdphi}
\dddphi+3H\ddphi-\frac{1}{a^2}\nabla^2\dphi+V''(\phib)\dphi+16\pi G\frac{\dot\phib}{H}V'(\phib)\dphi+(8\pi G)^2\frac{\dot\phib^2}{H^2}V(\phib)\dphi=0\,.
\end{equation}
Given the invariance of this equation under spatial translations and anticipating the need to canonically quantize the perturbations, we look for solutions of the form
\begin{equation}
\dphi(t,{\bf x})=\int\frac{d^3 k}{(2\pi)^3}\left[\dphi_k(t)a({\bf k})e^{i {\bf k\cdot x}}+\dphi_k^*(t)a^\dagger({\bf k})e^{-i {\bf k\cdot x}}\right]\,,
\end{equation}
where $k$ is the magnitude of the comoving momentum ${\bf k}$. In our conventions the creation and annihilation operators $a^\dagger({\bf k'})$ and $a({\bf k})$ satisfy
\begin{equation}
\left[a({\bf k}),a^\dagger({\bf k'})\right]=(2\pi)^3\delta({\bf k}+{\bf k'})\,.
\end{equation}
The modes $\dphi_k(t)$ satisfy equation~\eqref{eq:eomdphi} with $\nabla^2$ replaced by $-k^2$.  At early times, the gradients dominate over the terms involving the potential. The mode equation can then be solved in the WKB approximation, and we can canonical quantize the field to find
\begin{equation}\label{eq:dphiin}
\dphi_k(t)\to\frac{1}{a(t)\sqrt{2k}}\exp\left[-i k\int\limits_{t_*}^t\frac{dt'}{a(t')}\right]\,.
\end{equation}
Finally, we assume that the universe is in the Bunch-Davies vacuum,\footnote{Departures from a Bunch-Davies initial state have been considered in the context of trans-Planckian corrections to the perturbation spectrum~\cite{Brandenberger:1999sw,Easther:2002xe}, and would need to be substantial in order to undermine the conclusions reached here.} defined by
\begin{equation}
a({\bf k})\left|0\right\rangle=0\,.
\end{equation}

Now consider modes outside the horizon, with  $k/aH\ll1$.
In this limit the gauge invariant quantities $\mathcal{R}$ and $\zeta$ approach   constants. In our gauge\footnote{The careful reader may be worried about the appearance of $\dot\phib$ in the denominator of these expressions. This is an artifact of the definition of these quantities at linear order in perturbations:  they are only gauge invariant as long as $\dot\phib\neq 0$.}
\begin{equation}\label{eq:rdef}
\mathcal{R}_k=-H\frac{\dphi_k}{\dot\phib}\, , \qquad \zeta_k=\frac{\delta\rho_k}{3\dot\phib^2}\,.
\end{equation}
The perturbation to the energy density of the scalar field $\delta\rho_k$ is given by
\begin{equation}\label{eq:rhodef}
\delta\rho_k=\dot\phib\ddphi_k+V'(\phib)\dphi_k-\delta N_k\,\dot\phib^2\,,
\end{equation}
with
\begin{equation}\label{eq:dnphi}
\delta N_k=\frac{4\pi G \dot\phib}{H}\dphi_k\,.
\end{equation}
In the limit $k/aH\to 0$
\begin{equation}\label{eq:sol1}
\dphi_k(t)=-\mathcal{R}_k^{(o)}\frac{\dot\phib}{H}\,,
\end{equation}
is a solution of~\eqref{eq:eomdphi} for constant $\mathcal{R}_k^{(o)}$. For this solution one has $\mathcal{R}_k=\mathcal{R}_k^{(o)}$, and the superscript $(o)$ denotes the value outside the horizon.  With one solution known, we  obtain the second solution,
\begin{equation}\label{eq:sol2}
\dphi_k(t)=-\mathcal{C}_k\frac{\dot\phib}{H}\int\limits_{t_*}^t\;dt'\;\left(\frac{a(t_*)}{a(t')}\right)^3\frac{H(t')^2}{\dot\phib(t')^2}\,.
\end{equation}
During an inflationary period, the second term decays exponentially in time and $\mathcal{R}_k$ rapidly approaches the constant $\mathcal{R}_k^{(o)}$, and remains fixed until the mode re-enters the horizon.     In the post-inflationary phase, the second solution again decays, but like $1/t^2$ rather than exponentially.
 From these two equations, we see that
\begin{equation}
\mathcal{R}_k=\mathcal{R}_k^{(o)}+\mathcal{C}_k\int\limits_{t_*}^t\;dt'\;\left(\frac{a(t_*)}{a(t')}\right)^3\frac{H(t')^2}{\dot\phib(t')^2}\,.
\end{equation}
The solutions~\eqref{eq:sol1} and~\eqref{eq:sol2} are exact solutions of equation~\eqref{eq:eomdphi} in the limit of $k/aH\to 0$ and no assumption about the potential, or the existence of an inflationary phase is necessary.   Finally,   equations~\eqref{eq:rdef},~\eqref{eq:rhodef}, and~\eqref{eq:dnphi} together with equation~(\ref{eq:phibackground}) imply that for a single scalar field minimally coupled to gravity
\begin{equation}\label{eq:zetafromr}
\zeta_k=\mathcal{R}_k-\frac{\dot{\mathcal{R}}_k}{3H}\,,
\end{equation}
so that $\zeta_k$ and $\mathcal{R}_k$ become equal after horizon exit when the second mode has decayed.
 
In the preceding discussion, we recapitulated the usual treatment of
cosmological perturbations in scalar field theories, highlighting
two important points. First, we work directly with $\dphi_k$ which
remains finite and smooth even for $\dot{\phib} \rightarrow 0$, both
when manipulating the action (by refraining from substituting the
equation of motion for the $V''$ term) and in our choice of the
independent variable itself. The equations of motion are then both
easy to solve numerically and convenient to analyze analytically
with WKB methods. Second, we explicitly identify the decaying
solution for $\dphi_k$. One might expect it to be important for
modes whose wavelength is close to the post-inflationary horizon
size. Knowledge of the decaying solution allows us to check
that it is negligible even for these modes at sufficiently late
times. 


\section{Post-Inflationary Dynamics}

We now turn to the study of the post-inflationary period, during which the field undergoes damped oscillations around the minimum of its potential. For small enough amplitude, almost any potential is well approximated by the leading term in its Taylor expansion\footnote{Without loss of generality, we can stipulate that the minimum occurs at $\varphi=0$, and we are implicitly setting the vacuum energy to zero.}
\begin{equation}
V(\varphi)=\frac12 m^2\varphi^2\,.
\end{equation}
Our main interest is to understand when and on what scales the first nonlinearities arise. Including higher order terms  in the Taylor expansion turns on self-couplings of the scalar field and will necessarily modify the results. However, we expect that these interactions will generically speed up, rather than impede, the formation of non-linearities.  Neglecting all self-couplings is thus expected to provide the most conservative estimate of the duration of the phase of coherent oscillations.    For simplicity, we will assume that the potential is not only quadratic during the post-inflationary era but also during inflation. Nothing in our analysis depends directly on the detailed form of the perturbation spectrum at astrophysical scales, but we do tacitly assume that inflation ends near the GUT scale, so our analysis below would need minor modifications when applied to low scale inflation.

For our numerical calculations, we assume that the pivot scale
$k_*=0.002\;\text{Mpc}^{-1}$ exits the horizon $60$ $e$-folds before
the end of inflation, and we take $m=6.35\times 10^{-6}\, M_p$, to
be roughly consistent with the normalization of the power spectrum
given in~\cite{Komatsu:2010fb}. The precise value of $m$ does not
make a material difference to any of the statements below. However,
it is worth noting that if we do have a long matter dominated phase
$N$ and $m$ need to be solved self-consistently. 

\subsection{Background Dynamics}

 By construction $\epsilon=1$  when $\ddot{a}=0$, and sets the endpoint of inflation. At this instant
\begin{equation}
H_\text{end}=\frac{m}2\frac{\phib_\text{end}}{M_p}\,,
\end{equation}
where $M_p=1/\sqrt{8\pi G}$ is the reduced Planck mass.
To a good approximation $\phib_\text{end}\approx M_p$ and
\begin{equation}
H_\text{end}\approx\frac{m}2 \,.
\end{equation}
The Hubble parameter decreases rapidly thereafter so that $H\ll m/2$ during the oscillatory phase.  Thus a at late times we can solve  for $\phib$ and $H$ in the WKB approximation. At leading order we recover the  well known result~\cite{Starobinsky:1978S}
\begin{equation}
\phib(t)\approx\sqrt{6}M_p \frac{H}{m}\sin(m t +\Delta)\,,
\end{equation}
where $\Delta$ is some constant phase. With appropriate choice of the zero of time the expansion rate of the universe during this period is given by
\begin{equation}
H\approx H_m=\frac{2}{3t}\,.
\end{equation}
The index $m$ indicates that this is the expansion rate expected for a matter dominated universe.

\begin{figure*}[tb]
\centering
  \includegraphics[width=6.5in]{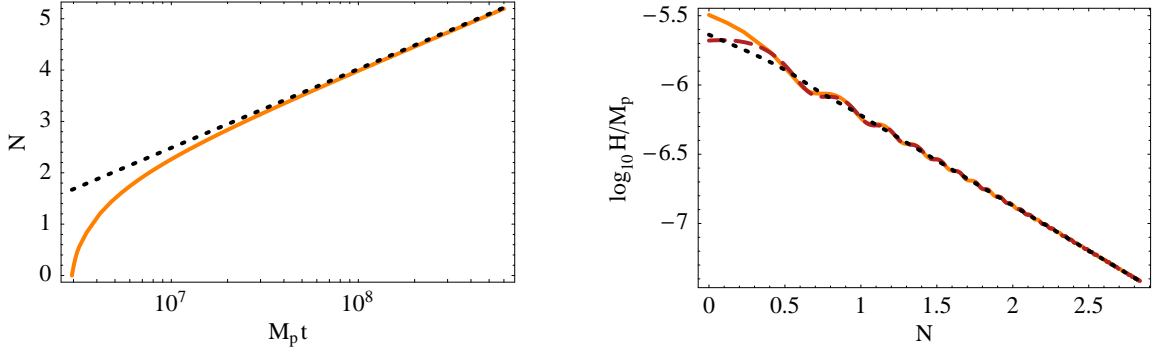}
  \caption{The left plot shows the number of $e$-folds after the end of inflation as a function of time. The right plot shows the Hubble constant as a function of the number of $e$-folds after the end of inflation. The orange lines represent the result of a numerical integration. The red curve with the longer dashes represent the analytic results of equation~\eqref{eq:bgsol}. The black line with the shorter dashes represents the leading order WKB approximation.}\label{fig:bg}
\end{figure*}

\begin{figure*}[ht]
\centering
  \includegraphics[width=6.5in]{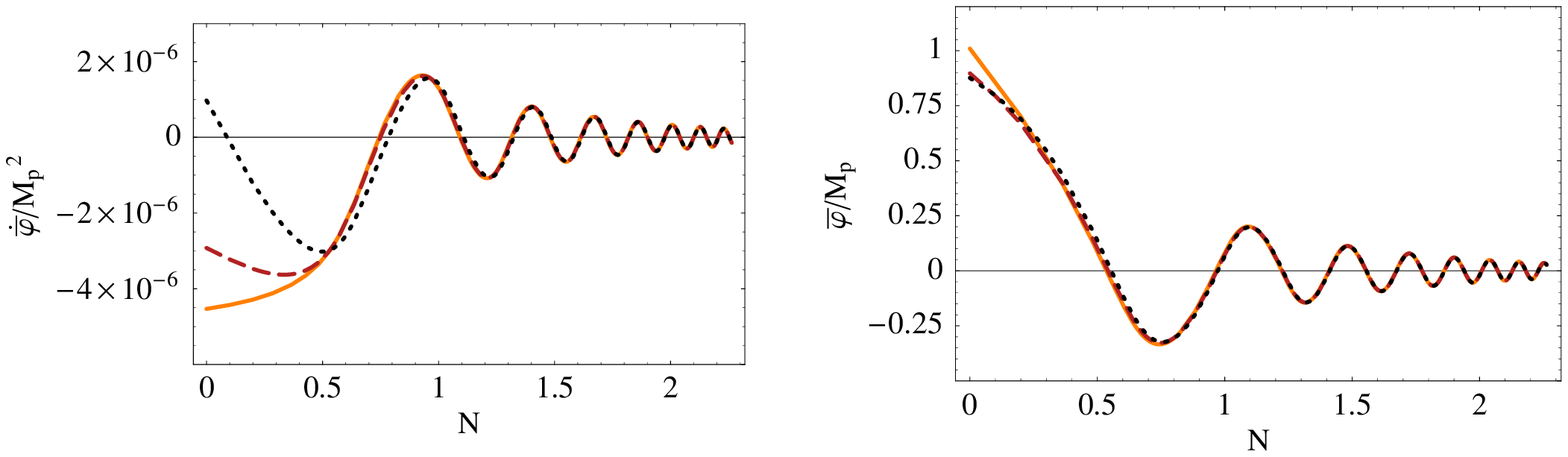}
  \caption{These plots show the evolution of the background field $\phib(t)$ and its derivative $\dot\phib(t)$ as a function of the number of $e$-folds after the end of inflation. The orange curves are the result of numerical integration. The red curve with the longer dashes represent the   analytic results of equation~\eqref{eq:bgsol}. The black line with the shorter dashes represents the leading order WKB approximation. }\label{fig:phiplot}
\end{figure*}

We use $\zeta_k$ as a gauge invariant measure of the growth of the
fractional energy density. It can be found by calculating
$\mathcal{R}$ and using equation~\eqref{eq:zetafromr}. We choose the
somewhat longer route using equations~\eqref{eq:rdef} together
with~\eqref{eq:rhodef} and~\eqref{eq:dnphi}, and verify that the two
agree. Since the denominator in $\zeta_k$ in
equation~\eqref{eq:rdef} is of order $H_m^2$, we need to know
$\delta\rho_k$ to order $H_m^2$. This requires us to compute the
background at second order in $H_m/m$,\footnote{ This may be
achieved either by going to the next order in the WKB approximation,
or alternatively by looking for a solution schematically of the form
$\phib=2\sqrt{2/3}M_p/mt+\sum
a_{kl}(mt)^{-l}\sin((2k-1)mt+\Delta)+\sum
b_{kl}(mt)^{-l}\cos((2k-1)mt+\Delta)$ and $H=2/3t+\sum
c_{kl}(mt)^{-l}\sin((2k)mt+\Delta)+\sum
d_{kl}(mt)^{-l}\cos(2kmt+\Delta)$ for positive integers $k$ and $l$
satisfying $l\geq 2$, and solving for the coefficients order by
order in an expansion in $1/mt$. This expansion is possibly
asymptotic, but we work at large $mt$ and truncate the series after
a few terms so issues of absolute convergence do not concern us. }
\begin{eqnarray}\label{eq:bgsol}
&&\hskip-.5cm\phib(t)=\sqrt{6}M_p \frac{H_m}{m}\left[\vphantom{\frac38}\sin(m t +\Delta)+\frac38\frac{H_m}{m}\cos(mt+\Delta)+\dots\right]\,,\\ 
&&\hskip-.5cm H(t)=H_m\left[1-\frac34\frac{H_m}{m}\sin\left(2(mt+\Delta)\right)+\dots\right]\,,
\end{eqnarray}
where the dots stand for corrections higher order in $H_m/m$.\footnote{The coefficient of the correction to the field depends on terms of third order in $H_m/m$ in our expansion for the background. Since we will not use these terms, we will not give them here.}   For the background scalar field the leading corrections have frequency $m$ and $3m$, and are suppressed by $(H_m/m)^2$. The leading correction to the expansion rate has frequency $2m$ and is also suppressed by $(H_m/m)^2$. The leading correction to the scale factor is of order $(H_m/m)^2$ (see also~\cite{Starobinsky:1978S}). We will thus take $a(t)=a_m(t)\propto t^{2/3}$. These results are shown together with the leading order result and the result of numerical integration in Figures~\ref{fig:bg} and~\ref{fig:phiplot}.

\subsection{Evolution of the perturbations}
We now turn to the post-inflationary evolution of the perturbations, and focus on the behavior of  non-relativistic modes, {\em i.e.\/} those with $k/am\ll1$  -- that is the {\em physical\/} wavenumber is much less than the inflaton mass. Since $a(t)$ is monotonic, once a mode becomes non-relativistic it remains so forever, in contrast to its transition across the horizon. Consequently, modes that leave the horizon during inflation are non-relativistic throughout the post-inflationary era.  A mode which has not left the horizon as inflation ends has $k/a_\text{end} > H_\text{end}$  and becomes non-relativistic after inflation is complete.  From  the previous section,   the dominant contribution for these modes in the limit $k/aH\to 0$ is given by~\eqref{eq:sol1}. Using~\eqref{eq:bgsol}, this becomes
\begin{equation}\label{eq:pertsol}
\dphi_k(t)=-\sqrt{6}M_p\mathcal{R}_k^{(\infty)}\left[\cos(mt+\Delta)-\frac32\frac{H_m}{m}\sin(mt+\Delta)+\frac38\frac{H_m}{m}\sin(3(mt+\Delta))\right]\,,
\end{equation}
where $\mathcal{R}_k^{(\infty)}$ is the value of $\mathcal{R}_k$ at times late enough so that not only $H_m/m\ll1$, but also $(k/a_mH_m)^2H_m/m\ll1$. We will see below how the second inequality arises.
When the inflaton oscillates around its minimum, the decaying solution is suppressed relative to this solution by a factor of $(H_m/m)^2$, and we neglect it. This implies that all modes eventually approach the dominant solution and become real whether they exit the horizon during the inflationary period or not.

\begin{figure}[ht]
\centering
\hskip-1.5cm\includegraphics[width=7in]{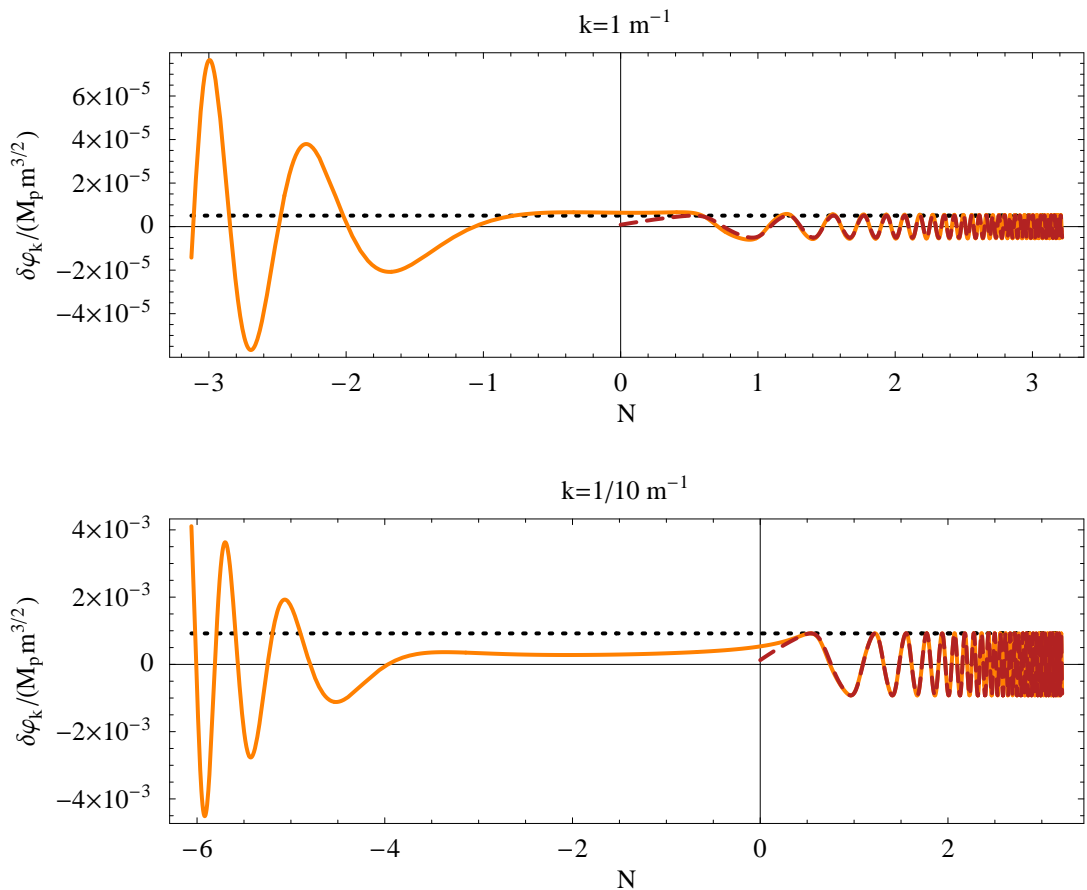}
  \caption{These plots show the evolution of the modes $\dphi_k(t)$ for different values of comoving momentum (in inverse meters) as a function of the number of $e$-folds after the end of inflation. The orange curves are the result of numerical integration. The red curve with the longer dashes represent the   analytic results of equation~\eqref{eq:pertsol}. The black line with the shorter dashes represents the leading order WKB approximation.  }\label{fig:dphiplot}
\end{figure}

For non-relativistic modes, we can calculate the momentum dependence to leading order in an expansion in $k/am$. For this purpose, it is sufficient to know the equation of motion for the perturbations~\eqref{eq:eomdphi} to linear order in our expansion in $H_m/m$. This implies that the last term in equation~\eqref{eq:eomdphi} is negligible, and the equation takes the form
\begin{equation}\label{eq:eomdphi2}
\dddphi_k+3H_m\ddphi_k+\frac{k^2}{a_m^2}\dphi_k+m^2\left[1+6\frac{H_m}{m}\sin(2(mt+\Delta))\right]\dphi_k=0\,.
\end{equation}
Calculating the leading correction due to $k/am$ to leading order in $H_m/m$, one finds
\begin{multline}
\dphi_k(t)=-\sqrt{6}M_p\mathcal{R}_k^{(\infty)}\left[\cos(mt+\Delta)-\frac32\frac{H_m}{m}\sin(mt+\Delta)+\frac38\frac{H_m}{m}\sin(3(mt+\Delta))\right.
\\\left.-\frac15\left(\frac{k}{a_mH_m}\right)^2\frac{H_m}{m}\sin(mt+\Delta)\right]\,.
\end{multline}
Notice that the momentum dependent piece decays like $t^{1/3}$ so that the correction  is always small at late times. The amplitude of these modes thus approaches a constant, $-\sqrt{6}M_p\mathcal{R}_k^{(\infty)} $. This solution together with a numerical solution of equation~\eqref{eq:eomdphi} is shown for different values of comoving momentum in Figure~\ref{fig:dphiplot}.
 
This solution immediately allows us to calculate $\mathcal{R}_k$ and $\zeta_k$. One finds
\begin{equation}
\mathcal{R}_k=\mathcal{R}_k^{(\infty)}\left[1-\frac{1}{5}\left(\frac{k}{a_mH_m}\right)^2\frac{H_m}{m}\tan(mt+\Delta)+\dots\right]\,.
\end{equation}
The dots indicate that we have dropped terms that are of the same order in $k/am$ but higher order in $H_m/m$, as well as terms higher order in both expansions.
At late times $\mathcal{R}_k$  approaches a constant, whether the modes are outside the horizon or not.  Note that while the late time limit $\mathcal{R}_k^{(\infty)}$ is well-defined, the approach to this limit is not monotonic, thanks to the presence of the  $\tan(mt+\Delta)$ term, ensuring that $\mathcal{R}_k$   acquires a ``spike'' whenever this tangent diverges.  The prefactor to this term scales like $t^{-1/3}$, so the spikes  get narrower, but are always present at this order in perturbation theory.  It is because of these divergences that we chose to work with $\delta\varphi_k$, which is a smooth and well-behaved function -- greatly facilitating both our analytical and numerical discussions.

Let us pause to understand the ``microscopic'' origin of the constancy of $\mathcal{R}_k$. On time scales long compared to $1/m$ but short compared to $1/H$, an inspection of equation~\eqref{eq:eomdphi} reveals that the modes are in parametric resonance, provided $k/a_m\lesssim\sqrt{3 H_m m}$, or equivalently $(k/aH_m)^2H_m/3m\lesssim 1$, as  already pointed out in~\cite{Nambu:1996gf,Jedamzik:2010dq}.  It turns out that this resonance actually yields  the solution with constant $\mathcal{R}_k$.     To substantiate this, consider  a mode that is observable in the CMB.  At early times, the mode is inside the horizon and $\dphi_k(t)$ oscillates with a frequency set by its momentum. The amplitude of $\dphi_k(t)$ decays like $1/a$, as in equation~\eqref{eq:dphiin}. Once the mode exits the horizon (long before the end of inflation), it freezes out and remains constant until the expansion rate drops below the mass. At this point $\dphi_k(t)$ becomes  oscillatory, even if it is still far outside the horizon. In the absence of the term driving the resonance identified already in~\cite{Nambu:1996gf,Jedamzik:2010dq}, the mode would decay like $1/t$ during this period, violating the constancy of $\mathcal{R}_k$  outside the horizon.

Parametric resonance is usually synonymous with exponential growth, but in this case the time-dependent  amplitude of the term driving the parametric resonance provides just enough amplification to keep $\mathcal{R}_k$ constant for {\em all\/} modes with $(k/a_mH_m)^2H_m/3m\lesssim 1$ whether inside or outside the horizon.  Thus this ``parametric resonance'' ensures that not only the background but also the perturbations behave very similar to the ones in a matter dominated universe, in which $\mathcal{R}_k$, or equivalently the Newtonian potential, is also constant both for modes inside and outside the horizon.

Now let us turn to $\zeta_k$. From a direct calculation, we find that
\begin{equation}
\zeta_k=\mathcal{R}_k^{(\infty)}\left[1+\frac{1}{15}\left(\frac{k}{a_mH_m}\right)^2\frac{1}{\cos^2(mt+\Delta)}+\dots\right]\,,
\end{equation}
where the dots again indicate terms that are higher order in the expansions in $H_m/m$, $k/am$.
This satisfies the relation~\eqref{eq:zetafromr} between $\zeta_k$ and $\mathcal{R}_k$. Further, the second term grows linearly in the scale factor, just as in a matter dominated universe.    One can make the correspondence even more precise by considering a time averaged version of $\zeta_k$ defined as
\begin{equation}
\overline{\zeta_k}\equiv\frac{\left\langle{\delta\rho_k}\right\rangle_T}{3\left\langle{\dot{\phib}^2}\right\rangle_T}\,,
\end{equation}
where the relevant quantities are averaged over some time $T$ much longer than $1/m$, but  short compared to the expansion rate of the universe. Using equations~\eqref{eq:rhodef},~\eqref{eq:dnphi} and~\eqref{eq:pertsol}, one finds
\begin{equation}
\overline{\zeta_k}=\mathcal{R}_k^{(\infty)}\left[1+\frac{2}{15}\left(\frac{k}{a_mH_m}\right)^2\right]\,,
\end{equation}
which is precisely what one expects for a matter dominated universe.
Alternatively, we can calculate the density contrast $\delta_k$
directly, where
\begin{equation}
\delta_k\equiv\frac{\delta\rho_k}{\overline\rho}\;, \qquad \overline\rho=\frac12\dot\phib^2+\frac12m^2\phib^2\,.
\end{equation}
It takes the form
\begin{equation}
\delta_k=\mathcal{R}_k^{(\infty)}\left[6\cos^2(mt+\Delta)+\frac{2}{5}\left(\frac{k}{a_mH_m}\right)^2\right]\,.
\end{equation}
The second term in brackets increases with time, so we expect the system to become nonlinear, and study this in detail in~\ref{s:nonlinear}.

We finish this subsection by computing  $\mathcal{R}_k^{(\infty)}$. For modes that exit the horizon before the end of inflation,   $\mathcal{R}_k^{(\infty)} = \mathcal{R}_k^{(o)}$. In general, $\mathcal{R}_k^{(o)}$ has to be calculated numerically, but for modes which  exit the horizon while the field is slowly rolling, the slow-roll approximation can be used and one has
\begin{equation}\label{eq:dr2sr}
\left|\mathcal{R}_k^{(\infty)}\right|^2=\frac{H^2(t_k)}{4\epsilon(t_k){M_p}^2}\frac{1}{k^3}\,,
\end{equation}
where $t_k$ denotes the time at which the mode $k$ exits the horizon.

We can also calculate this quantity for modes that never leave the horizon and  satisfy $k/aH\gg 1$ at all times. For these modes, as long as $k/a_m\gg \sqrt{3m H_m}$, the solution to equation~\eqref{eq:eomdphi} is well approximated by the WKB solution
\begin{equation}
\dphi_k(t)\to\frac{1}{a^{3/2}(t)\sqrt{2E(t)}}\exp\left[-i\int\limits_{t_*}^t dt' E(t')\right]\;\;\;\;\text{with}\;\;\;\;E(t)=\sqrt{\left(\frac{k}{a(t)}\right)^2+m^2}\,.
\end{equation}
Once $k/a_m$ becomes comparable to $\sqrt{3m H_m}$, parametric resonance sets in and the WKB solution is no longer valid. As in~\cite{Polarski:1992dq}, in this regime we look for a solution of the form
\begin{equation}
\dphi_k(t)=\frac{A(t)}{a^{3/2}(t)\sqrt{2m}}e^{-i(mt+\Delta)}+\frac{B(t)}{a^{3/2}(t)\sqrt{2m}}e^{i(mt+\Delta)}\,.
\end{equation}
Dropping terms suppressed by $k/am$ and introducing the independent variable 
\begin{equation}
x=\frac{k^2}{a^2H_m^2}\frac{H_m}{m}\,,
\end{equation}
the equation of motion for the perturbations of the scalar~\eqref{eq:eomdphi} turns into a coupled system of equations for $A$ and $B$\footnote{Dropping terms suppressed by $k/am$ is equivalent to treating $|\dot{A}/A|\ll m$ and $|\dot{B}/B|\ll m$ and working to leading order in the WKB approximation.}
\begin{eqnarray}
&&A'-iA+3\frac{B}{x}=0\,,\\
&&B'+iB+3\frac{A}{x}=0\,,
\end{eqnarray}
where the prime indicates a derivative with respect to $x$. This system can be solved by converting it into a second order differential equation for $A$. The initial conditions are chosen by matching to the WKB solution in the regime where both solutions are valid, {\it i.e.} for $\sqrt{mH}\ll k/a\ll m$. One finds
\begin{eqnarray}
&&A(x)=e^{ix}\left(1+\frac{9i}{2x}-\frac{9}{x^2}-\frac{15i}{2x^3}\right)\,,\\
&&B(x)=e^{ix}\left(\frac{3i}{2x}-\frac{6}{x^2}-\frac{15i}{2x^3}\right)\,.
\end{eqnarray}
At late times, when $x\ll 1$ 
\begin{equation}
\delta\varphi_k\to\frac{15}{\sqrt{2m}}i\frac{a^{9/2}m^3H_m^3}{k^6}\cos(mt+\Delta)\,.
\end{equation}
From this, we obtain $\mathcal{R}_k^{(\infty)}$
\begin{equation}\label{eq:dr2rel}
\left|\mathcal{R}_k^{(\infty)}\right|=\frac59\frac{(3m H_m)^3 a_m^{9/2}}{\sqrt{12 m}k^6}\;\;\;\;\text{or}\;\;\;\;\Delta_\mathcal{R}^2(k)=\frac{75 m^5 H_m^6}{8\pi^2M_p^2}\frac{a_m^9}{k^9}\,.
\end{equation}
Notice that $H_m^6a_m^9$ is a constant during this period, and we can evaluate it at any convenient time. The key result here is that the power spectrum decreases  like $\sim1/k^9$ for these modes.

\begin{figure*}[tb]
\centering
  \includegraphics[width=6.2in]{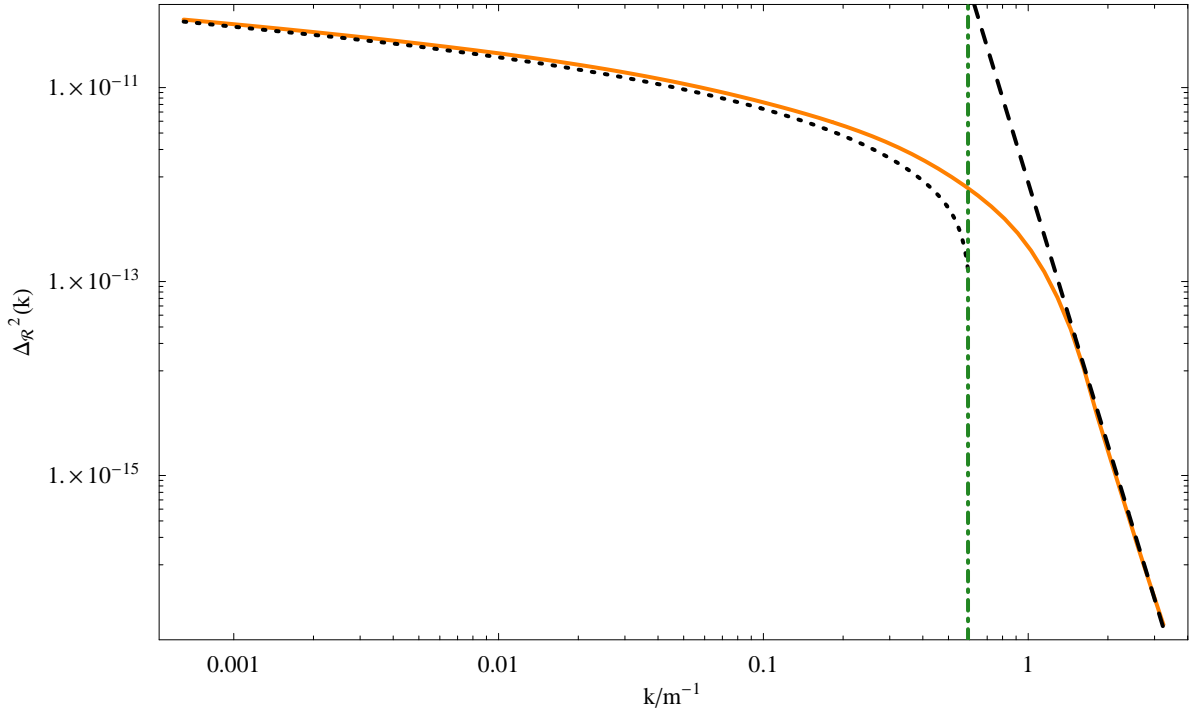}
  \caption{This plot shows the power spectrum $\Delta_\mathcal{R}^2(k)$ as a function of comoving momentum. The orange curve is the result of numerical integration.  The black line with the short dashes represents the result of the slow-roll approximation~\eqref{eq:dr2sr}. The black line with longer dashes represents the result~\eqref{eq:dr2rel}. Beyond the green vertical line modes no longer exit the horizon. As a consequence, one can no longer evaluate~\eqref{eq:dr2sr}. }\label{fig:dr2}
\end{figure*}

We now know the power spectrum both for modes that are highly non-relativistic near the end of inflation, and for modes that are highly relativistic at that time. For the transition region in which modes are neither highly relativistic nor highly non-relativistic at the end of inflation, we will rely on a numerical evaluation. The result is shown in Figure~\ref{fig:dr2}. At either extreme our result recovers the slow-roll approximation for the power spectrum, and our \eqref{eq:dr2rel}. We find excellent agreement between our analytic and numerical results in their appropriate regimes of validity.

\subsection{Growth of structures and the breakdown of linear perturbation theory}\label{s:nonlinear}

For modes inside the horizon, $\zeta_k$  grows linearly with the scale factor, indicating that perturbation theory will eventually break down. Let us now make this more precise. 

To ensure the validity of perturbation theory, $\delta N$ must remain small compared to the unperturbed value of the lapse, {\em i.e.} unity. To assess this, it is convenient to look at the variance
\begin{equation}
\left\langle\delta N(t,0)^2\right\rangle=9\cos^4(mt+\Delta)\int\frac{dk}{k}\Delta_\mathcal{R}^2(k)\,.
\end{equation}
The amplitude of this oscillatory function is constant in time, and thus does not signal a breakdown of perturbation theory at late times. The integral is sensitive to the contributions of modes with very long wavelengths and naively diverges for a power spectrum that is scale invariant in the limit $k\to0$. However, we only expect the spectrum to be scale invariant for modes that exited the horizon during the inflationary period, and we expect the power in modes with even longer wavelengths to be suppressed due to causality, making the integral convergent. Under these circumstances, the integral is sensitive to the total number of $e$-folds of inflation. Since the amplitude of the observed perturbations is rather small, this does not pose a serious bound on the total number of $e$-folds, and the linear treatment is valid.

\begin{figure*}[tb]
\centering
  \includegraphics[width=6.2in]{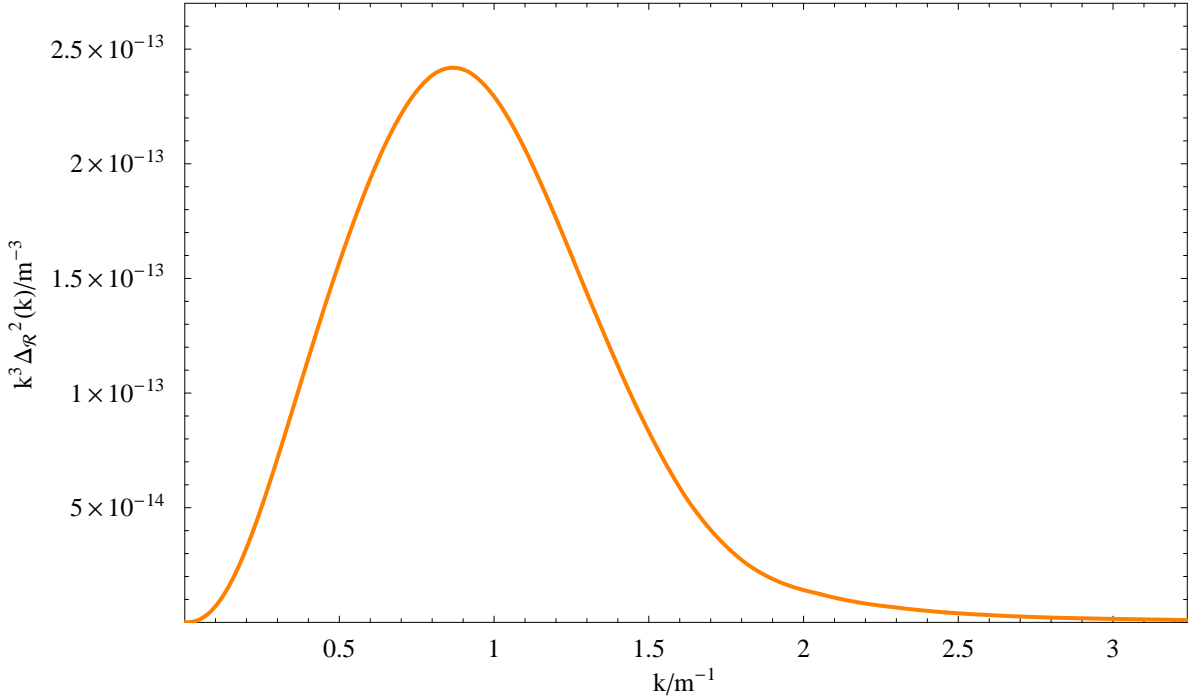}
  \caption{This plot shows the result of a numerical calculation of the integrand in equation~\eqref{eq:k3dr2}, $k^3\Delta_\mathcal{R}^2(k)$, as a function of comoving momentum. For small $k$ the curve increases like $k^3$, for large $k$ it decays like $1/k^6$. The integrand peaks for values of $k$ comparable to the comoving horizon size at the end of inflation.}\label{fig:k3dr2}
\end{figure*}

Next, let us look at $N^i$. To assess its importance, we compare the contribution to the extrinsic curvature of the spatial slices due to $N^i$ to the contribution from the background, and require that the ratio remain small. This translates into
\begin{equation}
\frac{1}{H^2}\left\langle(\partial_i N^i(t,0))^2\right\rangle\ll 1\,.
\end{equation}
Using~\eqref{eq:consni}, together with the results in the last subsection, one finds
\begin{equation}\label{eq:k3dr2}
\frac{1}{H^2}\left\langle(\partial_i N^i(t,0))^2\right\rangle=\frac{9}{25}\left(\frac{1}{a_mH_m}\right)^4\int dk\;k^3\Delta_\mathcal{R}^2(k)\,.
\end{equation}
The integrand is shown in Figure~\ref{fig:k3dr2}.
As one would expect, the integral receives its largest contribution from the modes whose physical momentum at the end of inflation is close to the Hubble constant at that time. These are modes in the transition region where the power spectrum changes from nearly scale invariant to decaying like $1/k^9$. We see that this quantity grows like $a^2$ during the post-inflationary era, and this term is the one that fixes the onset of nonlinearity. The integral has excellent convergence properties both for small and large momenta, and can easily be performed numerically. We find\footnote{This formula is only valid after the relevant modes have approached the solution with $\mathcal{R}_k$. This happens about 4 $e$-folds after the end of inflation.}
\begin{equation}
\frac{1}{H^2}\left\langle(\partial_i N^i(t,0))^2\right\rangle\approx1.4\times 10^{-12}\; e^{2N}\,,
\end{equation}
where $N$ denotes the number of $e$-folds after the end of
inflation.
We conclude that the onset of
non-linearity occurs roughly $13$ $e$-folds after the end of
inflation. Mapping between $H(t)$ and $a(t)$, we see that
non-linearities develop when $H\approx3\times 10^{-8}\;H_\text{end}$
or $H\approx 10^{-14}M_p$. If the universe thermalized
instantaneously at this point, we would have a reheat temperature
\begin{equation}
T_\text{rh}\approx\frac{4\times 10^{11}\;\text{GeV}}{\mathcal{N}_\text{eff}^{1/4}}\,,
\end{equation}
where $\mathcal{N}_\text{eff}$ is the effective number of relativistic degrees of freedom.

It may be interesting to point out that variance of the density contrast $\delta(t,{\bf x})=\delta\rho(t,{\bf x})/\overline{\rho}$ approaches unity around the same time. At late times it is given by
\begin{eqnarray}
&&\nonumber\left\langle(\delta(t,0))^2\right\rangle=36\cos^4(mt+\Delta)\int dk\;\Delta_\mathcal{R}^2(k)\\
&&\hskip2cm+\frac{24}{5}\cos^2(mt+\Delta)\left(\frac{1}{a_mH_m}\right)^2\int dk\;k\Delta_\mathcal{R}^2(k)\nonumber\\
&&\hskip2cm+\frac{4}{25}\left(\frac{1}{a_mH_m}\right)^4\int dk\;k^3\Delta_\mathcal{R}^2(k)\,.
\end{eqnarray}
The first term is small for any realistic total number of $e$-folds of inflation. The third term becomes large around $13$ $e$-folds after the end of inflation, while the second term grows more slowly.

Finally, let us consider the scalar perturbations. Two things have to be satisfied to justify treating the scalar field perturbations to linear order. First, their variance must remain sub-Planckian. At times late enough for the modes to have become real and to have approached the solution that oscillates with constant amplitude, this variance is given by
\begin{equation}
\left\langle\dphi(t,0)^2\right\rangle=6M_p^2\cos^2(mt+\Delta)\int\frac{dk}{k}\Delta_\mathcal{R}^2(k)\,.
\end{equation}
The amplitude of this oscillatory function is again independent of time, and remains small at late times.

Second, the scalar field perturbations should be small compared to the scalar field background to ensure that the energy density stored in the background dominates over that in the perturbations so that the perturbations do not affect the background geometry. This requires $\left\langle\dphi(t,0)^2\right\rangle\ll\overline\varphi^2(t)$, which is valid as long as
\begin{equation}
\frac{m^2}{H_m^2}\int\frac{dk}{k}\Delta_\mathcal{R}^2(k)\approx  2.9\times 10^{-8}e^{3N}\ll 1\,,
\end{equation}
where $N$ denotes the number of $e$-folds after the end of inflation. This becomes of order unity around six $e$-folds after the end of inflation and suggests a breakdown of our perturbation theory even earlier than we had found by looking at the metric perturbations. As was argued in~\cite{Polarski:1992dq}, however, this effect could be resummed so that the evolution can be studied analytically even after this happens and the system truly becomes non-linear only once the density contrast approaches unity.\footnote{We are very grateful to Alexei Starobinsky for pointing this out to us.}

\section{Discussion}

We have given a full and careful account of the evolution of the universe during a phase of ``coherent oscillations'' following inflation.  We have worked in variables that are manifestly finite, and given reliable analytic approximations for the individual modes' evolution.  As is well known, the background solution is pressure-free when averaged on time scales long compared to the oscillation time of the fields. All modes grow linearly with the scale factor,  matching our expectations for a dust-dominated universe with vanishing pressure. Modes whose natural frequency is much lower than the frequency of the scalar field oscillations during the end of inflation possess a nearly scale invariant power spectrum. For modes whose frequency during the end of inflation exceeds that of the background field oscillations, the power spectrum decreases rapidly with increasing frequency.

The breakdown of the analogy between the coherent oscillations of the inflaton and pressureless dust  reflects the presence of two intrinsic time scales in our dynamical system, namely the Hubble time and the oscillation period for the scalar field condensate. Conversely,  a dust-dominated universe has only the Hubble time. Consequently, the perturbations of the scalar field match those of dust only when we can ``integrate out'' the more rapid scalar field oscillations.  Interestingly,  the ``long modes'' for which the pressureless dust approximation is reliable can be shown to undergo parametric resonance while they are outside the horizon~\cite{Jedamzik:2010dq}, and this resonance is the microphysical origin of the constancy of $\mathcal{R}_k$, as we show via an explicit calculation.
 
Phenomenologically, this distinction between ``long'' and ``short'' modes  picks out a unique scale  in the post-inflationary universe.  At the end of inflation, modes which are approximately horizon-sized have a frequency similar to the oscillations of the coherent inflaton background. Solutions for modes well inside the horizon at the end of inflation decay in amplitude until they are redshifted to lower frequencies by the expansion of the universe.  Conversely, modes that left the horizon well before the end of inflation are frozen until they re-enter the horizon.  Consequently, modes that are roughly horizon sized at the end of inflation undergo  significantly more growth than other modes and will be the first for which  nonlinear effects become important, and the breakdown of the coherent oscillations takes place on length scales fixed by the post-inflationary horizon size.

Physically, the breakdown of coherence in the oscillating field implies that while we can assume a very long period of matter domination following inflation, the universe can expand by at most  a factor $\sim e^{13}$ or so during the coherent oscillation phase. Beyond this point,  $\delta \rho/\rho \sim 1$, and the field is no longer homogeneous. Once this breakdown occurs, we need to move beyond first order perturbation theory, raising the prospect that even the simplest post-inflationary universe can have highly nontrivial phenomenology. This could include gravitational wave production \cite{Assadullahi:2009nf,Assadullahi:2009jc,Jedamzik:2010hq}, or even the formation of primordial black holes (e.g. \cite{1985MNRAS.215..575K}) if the overdensities grow without limit. In the latter case, the primordial black holes will be very small, since their mass will be set by the mass-energy inside a region whose size is equal to the comoving length scale at the end of inflation.  These black holes would thus decay quickly via Hawking radiation, reheating the universe (e.g. \cite{GarciaBellido:1996qt}) via ``non-perturbative'' gravitational effects. Moreover, in addition to any gravitational waves generated via the nonlinear evolution of the field, gravitons Hawking radiated as the black holes decay would generate a background of very high-frequency gravitational waves \cite{Anantua:2008am}.

More generally, this analysis reinforces the importance of carefully exploring the physical processes which  govern the universe between inflationary and nucleosynthesis scales. Models with parametric resonance are well known for their rich phenomenology and possible gravitational wave background (e.g. \cite{Easther:2006gt,Easther:2006vd,Easther:2007vj,GarciaBellido:2007af,Dufaux:2007pt}, but it is  becoming increasingly clear that   even the simplest potentials  can have nontrivial post-inflationary dynamics.  We also wish to understand the overall thermal history of the post-inflationary epoch in order to accurately match present-day astrophysical scales to their counterparts during the inflationary era \cite{Liddle:2003as}.   Consequently, while this model is almost as old as inflation itself \cite{Linde:1983gd}, its phenomenology contains a number of  important open questions. The analysis here provides a thorough and complete understanding of the post-inflationary oscillations, and we intend to pursue the issues raised here in future work.

\section{Acknowledgments}

We thank Peter Adshead and Mustafa Amin and in particular Alexei Starobinsky for valuable discussions, and Jerome Martin and David Wands for helpful comments on the draft.
The authors are partly supported by the Department of Energy (DE-FG02-92ER-40704), and RE and RF receive support from the NSF (CAREER-PHY-0747868).

\end{document}